\documentclass[useAMS,usenatbib]{mn2e}
\usepackage{graphicx}

\title[Binarity of Cepheids WW~Car and FN~Vel]
{Discovery of blue companions to two southern Cepheids: WW~Car and FN~Vel}
\author[V.~Kovtyukh et al.]
{V.~Kovtyukh$^{1,2}$\thanks{E-mail: val@deneb1.odessa.ua},
L.~Szabados$^{3}$,
F.~Chekhonadskikh$^{1,2}$,
B.~Lemasle$^{4}$,
and S.~Belik$^{1,2}$\\
$^{1}$Astronomical Observatory, Odessa National University,
       Shevchenko Park, 650014, Odessa, Ukraine\\
$^{2}$Isaac Newton Institute of Chile, Odessa branch,
       Shevchenko Park, 650014, Odessa, Ukraine\\
$^{3}$Konkoly Observatory, Research Centre for Astronomy and Earth
Sciences, Hungarian Academy of Sciences, H-1121 Budapest, \\
Konkoly Thege Mikl\'os \'ut 15-17, Hungary\\
$^{4}$Anton Pannekoek Astronomical Institute, Science Park 904, 
P.O. Box 94249, 1090 GE Amsterdam, The Netherlands}

\begin{document}

\date{Accepted 2015 February 4. Received 2015 February 3; in original form 2015 January 19}
\pagerange{\pageref{firstpage}--\pageref{lastpage}}
\pubyear{2015}

\maketitle

\label{firstpage}

\begin{abstract}
A large number of high-dispersion spectra of classical Cepheids were
obtained in the region of the Ca\,II\,H+K spectral lines. The analysis
of these spectra allowed us to detect the presence of a strong Balmer
line, H$\epsilon$, for several Cepheids, interpreted as the signature 
of a blue companion: the presence of a sufficiently bright blue 
companion to the Cepheid results in a discernible strengthening of 
the Ca\,II\,H + H$\epsilon$ line relative to the Ca\,II\,K line.
We investigated 103 Cepheids, including those with known hot 
companions (B5-B6 main-sequence stars) in order to test the method.
We could confirm the presence of a
companion to WW Car and FN Vel
(the existence of the former was only suspected before)
and we found that these companions are blue hot stars.
The method remains efficient when the orbital 
velocity changes in a binary system cannot be revealed and 
other methods of binarity detection are not efficient.

\end{abstract}

\section{Introduction}

Classical Cepheids are radially pulsating F-G supergiants. 
Their regular variability makes them ideal standard candles
in establishing the cosmic distance scale via the
period-luminosity ($P$-$L$) relationship. The calibration
of this relationship has a century-long history, and there
is still need for improving the zero-point \citep{FM10}.

Cepheids which are members of binary systems can be suitable
calibrators only if the luminosity of the Cepheid component can
be disentangled from the luminosity of the companion star: if 
the companion remains unrevealed, its photometric effect can 
falsify the luminosity determination of the Cepheid. Because 
the incidence of binaries exceeds 50 per cent among classical 
Cepheids \citep{Sz03a}, studying the binarity of individual 
Cepheids is an important task which is impeded by the fact 
that the companions are usually much fainter than the 
supergiant Cepheids.

Companions to Cepheids can be discovered by all conventional
methods used for revealing binarity involving spectroscopy,
photometry, and astrometry. There are some specific photometric
methods only applicable for Cepheids summarized by \citet{Sz03b}.

Because most of the detectable companions are early type
main-sequence (or slightly more evolved) stars whose flux
dominates the ultraviolet part of the binary spectrum, UV 
spectroscopy with the {\it IUE} satellite was especially 
successful in disclosing the binarity of Cepheids \citep{E92}. 
In the absence of UV spectra, there is a complementary method based on
a specific portion of the optical spectrum for detecting blue 
secondary stars efficiently.

\begin{figure}
\resizebox{\hsize}{!}
{\includegraphics{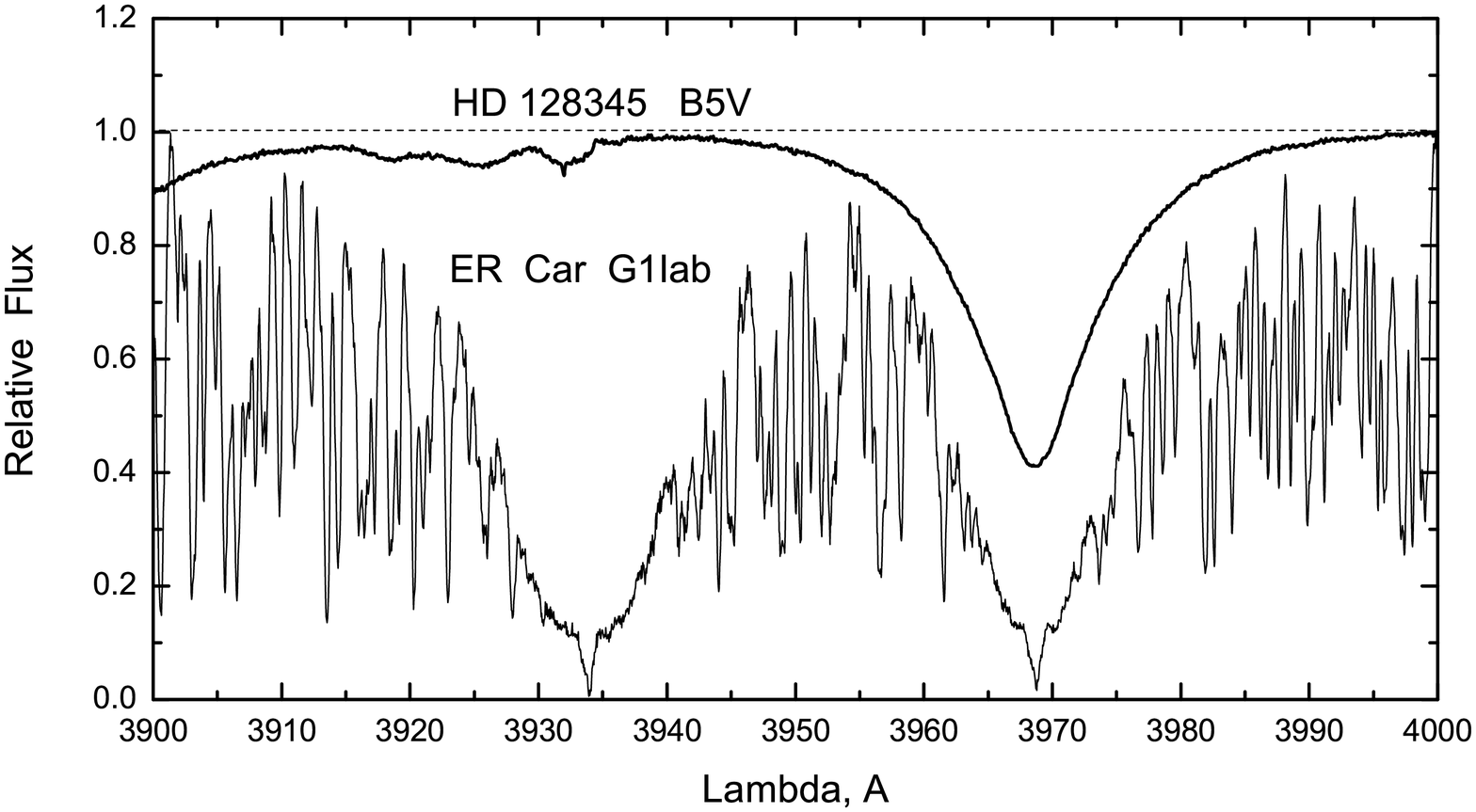}}
\caption[]{
Spectra in the region of the strong Ca\,II\,H\,+\,K spectral lines:
a typical Cepheid without bright companion (lower spectrum);
a main-sequence B-star (upper spectrum). The narrow absorption
features near the line cores are of interstellar origin.}
\label{ewr4}
\end{figure}

This method, referred to as the calcium-line method, is based
on the spectrophotometry of the 3900--4000~\AA\ part of the
visible region. This interval covers the Ca\,II\,H (3968.47~\AA) 
and the Ca\,II\,K (3933.68~\AA) lines as well as the H$\epsilon$ 
(3970.07~\AA) line of the hydrogen Balmer series.
The Ca\,II\,H and K lines in spectra of normal Cepheids without 
bright companions have practically equal depths (or residual 
flux). (We refer to the line profiles of Cepheids not accounting 
for narrow overlapping interstellar lines).
If, however, a hot companion is present, the H$\epsilon$ line 
from the secondary star is superimposed on the Ca\,II\,H line 
and the resulting blend of these two lines is deeper than the
Ca\,II\,K line (see Fig.~\ref{ewr4}). This technique of
`intensity reversal' was first used by Miller \& Preston
(\citeyear{MP64a,MP64b}) for Cepheids. Later on, \citet{E85}
showed that this method can be applied for detecting
companions hotter than A3V spectral type stars.

In the last three decades this method was neglected. In view
of the fact that there is no dedicated space mission for
obtaining UV spectra, we decided to perform an optical
spectroscopic survey of a large number of Cepheids to reveal
hot companions, i.e. new binaries.

\section{Observations}

The region of the Ca\,II\,H\,+\,K lines was studied in the spectra 
of all 103 Cepheids listed in the paper by \citet{Letal11}. 
The list of the target Cepheids is given in their Table~1.

High signal-to-noise spectra were obtained in the period 
25~March -- 1~April 2010 using the 2.2\,m MPG telescope and 
FEROS spectrograph at ESO La Silla. The spectra cover a continuous 
wavelength range from 3800 to 8700~\AA\ with a resolving power of 
about 48000. Typical maximum signal-to noise (S/N) values (per 
pixel) for the spectra are in excess of 150. Each night we observed 
a broad-lined B star with a S/N exceeding that of the programme 
stars to enable cancellation of telluric lines where necessary.

We used \textsc{iraf} to perform CCD processing, scattered light
subtraction, and echelle order extraction. For these spectra two
extractions were done, one uses a zero-order (i.e., the mean)
normalization of the flat field which removes the blaze from the
extracted spectra. The second one uses a high-order polynomial to
normalize the flat-field which leaves the blaze function in the
extracted spectrum.
The latter spectrum reflects more accurately the true counts
along the orders. A Windows based graphical package (ASP) developed 
by R. Earle Luck was used to process the blaze removed spectra.
This included Beer's law removal of telluric lines, smoothing
with a fast Fourier transform procedure, continuum normalization,
and wavelength calibration using template spectra.
Echelle orders show significant S/N variations from edge
to maximum due to blaze efficiency. To maximize the S/N in
these spectra we have co-added the order overlap region using
as weights the counts from the second data extraction.
The co-added spectra were then inspected and the continua
sometimes modified by minor amounts in the overlap regions.

Anomalous profiles of the Ca\,II\,H\,+\,K lines, indicative 
of the presence of a hot companion, were found only in ten of 
these spectra. The data on these Cepheids are listed in Table~1. 
The spectra of Cepheids were obtained predominantly at maximum 
brightness, in order to reach the highest S/N and to perform a 
full abundance analysis \citep{Letal11}. However, the optimal 
way to search for companions is to use spectra of Cepheids at 
minimum brightness when the contribution of possible companions 
to the spectra of the target Cepheid is the highest. Therefore 
we cannot exclude that a few hot companions remain unnoticed. 
In the case of known binary Cepheids, not all spectra show 
anomalous behaviour of the Ca\,II\,H\,+\,K lines, either because 
the companion is too faint, or it is of late spectral type.

\section{Effect of a blue companion on the 
C\lowercase{a}\,II\,H\,+\,K lines}
\subsection{Cepheids with known blue companion}
\label{knowncompanions}

\begin{figure*}
\resizebox{\hsize}{!}
{\includegraphics{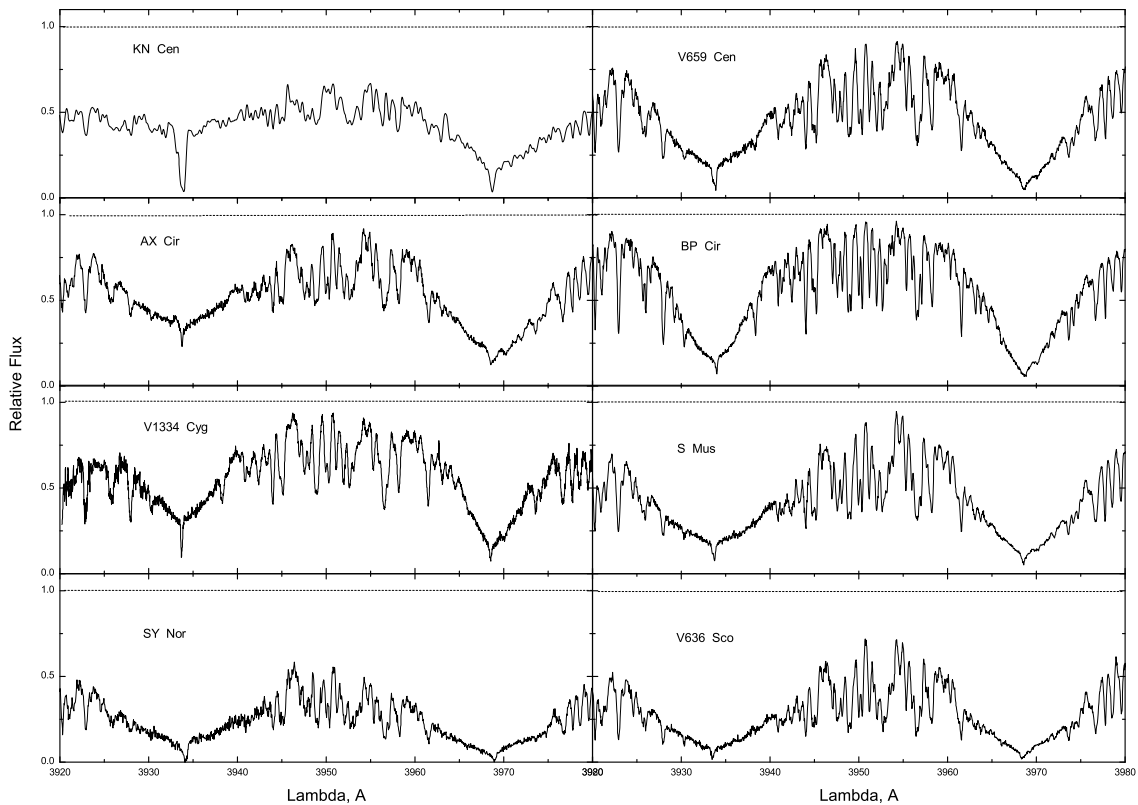}}
\caption[]{Spectra of 8 classical Cepheids with known 
B type companion in the region of the Ca\,II\,H+K lines.}
\label{fig2}
\end{figure*}

Cepheids with known blue companions have been selected
from the list of Galactic Cepheids in binary systems
(http://www.konkoly.hu/CEP/nagytab3.html). These target Cepheids
are listed in the upper section of Table~\ref{binceplist}
whose subsequent columns give the name of the Cepheid, the
pulsation period (in days), the mean $V$ brightness, the spectral type of
the Cepheid, the ratio of the residual fluxes defined below, and the spectral
type of the blue companion. The relevant part of the observed 
spectra is shown in Fig.~\ref{fig2}. It is clearly seen that 
the blend of the Ca\,II\,H and H$\epsilon$ lines is stronger 
than the Ca\,II\,K line.

To test the presence of a possible hot companion, we used 
the ratio of residual fluxes, $R_{\rm KH} = r_{\lambda}({\rm K})/r_{\lambda}({\rm H})$.
A typical value of the ratio for Cepheids without hot companions is 
$R_{\rm KH} = 1.00 \pm 0.03$.
As expected, the ratios for the stars with known hot companions 
show larger values ranging from 1.29 to 2.09 (see Table~1).

\begin{table}
\caption{Cepheids with anomalous profiles of
the Ca\,II\,H\,+\,K lines}
\begin{center}
\begin{tabular}{l@{\hskip0mm}r@{\hskip2mm}r@{\hskip2mm}l@{\hskip0mm}c@{\hskip1mm}c}
\hline
Cepheid & $P$ (d) & $\langle V \rangle $ & Sp(Cep) &$R_{\rm KH}$& Sp(comp)\\
\hline
{\it known binaries}: & & & &            &\\
KN Cen    & 34.0296 &  9.87 & G8Iab     & 2.09 &B6.0\,V \\
V659 Cen   & 5.6218 &  6.60 & F6/F7Ib   & 1.61 &B6.0\,V \\
AX Cir     & 5.2733 &  5.88 & F8II      & 1.67 &B6.0\,V \\
BP Cir     & 2.3984 &  7.56 & F2/F3II   & 1.27 &B6.0\,V \\
V1334 Cyg  & 3.3330 &  5.87 & F1II      & 1.86 &B7.0\,V \\
S Mus      & 9.6599 &  6.12 & F6Ib      & 1.44 &B3.5\,V \\
SY Nor     & 12.6457&  9.51 & F9Ib      & 1.92 &B4.5\,V \\
V636 Sco   & 6.7968 &  6.65 & F7/F8Ib/II& 1.29 &B9.5\,V \\
\hline
{\it new binaries}: & & & &         & \\
WW Car     & 4.6768 & 9.75 &F9II    & 1.35  & ? \\
FN Vel     & 5.3242 &10.29 &F9II    & 1.42  & ? \\
\hline
\end{tabular}
\end{center}
\label{binceplist}
\end{table}

\subsection{Cepheids with newly revealed blue companion}
\label{newcompanions}

\begin{figure}
\resizebox{\hsize}{!}
{\includegraphics{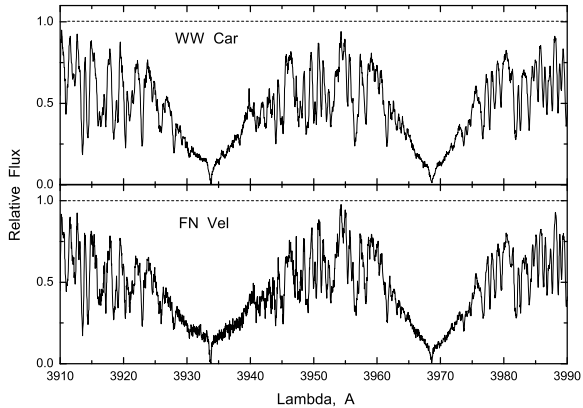}}
\caption[]{
Spectra of two new binary classical Cepheids WW~Car and FN~Vel.
It is noticeable, that the Ca\,II\,H + H$\epsilon$ line
is enhanced relative to the Ca\,II\,K line.}
\label{fig3}
\end{figure}

A similar intensity reversal was searched for in the
spectra of other Cepheids, and it was found that WW~Car
and FN~Vel have a formerly unrevealed blue companion.
Indeed, the r$_{\lambda}$(K)/r$_{\lambda}$(H) ratio has 
a value of 1.35 in the case of WW~Car and of 1.42 in the 
case of FN~Vel, well above 1.00. The basic data on these 
two Cepheid variables are listed in Table~\ref{log2} and 
their spectra near the Ca\,II\,H+K lines are shown in 
Fig.~\ref{fig3}.\\



\begin{table*}
\caption{New binary Cepheids with blue companions. 
The accurate value of the pulsation period (col.~2) is from 
Sect.~\ref{discussion}. The Julian Date of the spectral 
observation (col.~7) and the corresponding phase (col.~8) 
have been calculated from the newly determined ephemerides. 
The mean $V$ brightness (col.~3) and $B-V$ colour index 
(col.~4) in the Johnson photometric system are from 
\citet{Betal00}; the colour excess, $E(B-V)$ (col.~5) and 
the atmospheric iron content [Fe/H] (col.~6) are both from 
\citet{Letal11}.}
\begin{center}
\begin{tabular}{lrcrcrcrrr}
\hline

Cepheid & Period & $\langle V \rangle$ & $\langle B-V \rangle$ & $E(B-V)$ &
[Fe/H] & JD & phase & Exp. & S/N \\
   & (day) & (mag) & (mag) & (mag) &  & 2\,400\,000+ &  & (sec) & \\
\hline
WW Car & 4.676818 &  9.748 & 0.899& 0.379 & $-$0.07& 55282.603 & 0.849 & 1200 & 218 \\
FN Vel & 5.324170 & 10.292 & 1.186& 0.588 &  0.06  & 55283.591 & 0.224 & 2100 & 251 \\
\hline
\end{tabular}
\end{center}
\label{log2}
\end{table*}

\section{Discussion}
\label{discussion}

All information on both WW~Car and FN~Vel available in the
literature was collected for two main reasons, viz.:\\
- to find additional evidence of binarity;\\
- to determine the accurate value of the pulsation period
in order to calculate the phase of the spectral observation.

The updated period was determined by the $O-C$ diagram 
method \citep{S05}. The newly determined ephemerides can be 
used when planning any future observations of these Cepheids.

All published photometric observations of WW~Car and FN~Vel
were re-analysed in a homogeneous manner to determine
seasonal moments of the chosen light curve feature. The
relevant data listed in Tables~\ref{tab-wwcar-oc} and
\ref{tab-fnvel-oc}, respectively, are as follows:\\
\noindent - Column~1: the heliocentric moment of the selected light curve 
  feature (moment of maximum brightness);\\
\noindent - Col.~2: the epoch number, $E$, as calculated from 
  equations~(\ref{wwcar-ephemeris}) and (\ref{fnvel-ephemeris}), 
  respectively:
\vspace{-1mm}
\begin{equation}
C = 2\,453\,047.7725 + 4.676\,818{\times}E
\label{wwcar-ephemeris}
\end{equation}
\vspace{-3mm}
$\phantom{mmmmm}\pm0.0032\phantom{}\pm0.000\,002$
\begin{equation}
C = 2\,453\,775.6587 + 5.324\,170{\times}E
\label{fnvel-ephemeris}
\end{equation}
\vspace{-3mm}
$\phantom{mmmmm}\pm0.0022\phantom{}\pm0.000\,005$\\
\noindent (These ephemerides have been obtained by the weighted
linear least squares fit to the $O-C$ differences for both
Cepheids);\\
\noindent - Col.~3: the corresponding $O-C$ value;\\
\noindent - Col.~4: the weight assigned to the $O-C$ value (1, 2, 
or 3 depending on the quality of the light curve leading to
the given difference);\\
\noindent - Col.~5: the source of the data.

\subsection{WW~Carinae}
\label{sub-wwcar}

The variability of WW~Carinae was discovered by Henrietta
Leavitt \citep{P06}. The period determined somewhat later
by Arville Walker \citep{P12}, $P = 4.676$~d, is correct but 
no type of variability was assigned to the star. \citet{Sz26} 
published a series of photographic observations covering the 
years 1924-1926. In his paper, WW~Car is already considered as 
a Cepheid. The available photoelectric and more recent CCD 
photometric data involve those by \citet{Wetal58}, \citet{I61}, 
\citet{P76}, \citet{P02}, \citet{B08}, as well as the observations 
by the {\it Hipparcos} satellite \citep{ESA97}, and Optical 
Monitoring Camera (OMC) on board the {\it INTEGRAL} space probe. 
The most recent photometric data are accessible on the the AAVSO web 
page\footnote{http://www.aavso.org/apps/webobs/results/?star=WW+CAR}.
From multi-colour photometry, \citet{MF80} suspected the existence 
of a blue companion to WW~Car.\\

\begin{figure}
{\includegraphics[height=40mm, angle=0]{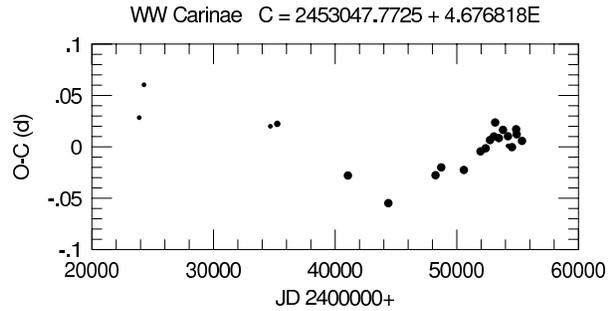}}
\caption[]{
$O-C$ diagram for WW~Car.}
\label{fig-wwcar-oc}
\end{figure}

The moment of maximum brightness was determined for each data
series from the annual phase curves. When constructing the $O-C$
diagram, the $O-C$ values for the moments of maximum 
brightness of WW~Car have been obtained from the ephemeris 
\ref{wwcar-ephemeris}, and are listed in Table~\ref{tab-wwcar-oc}. 
The $O-C$ diagram in Fig.~\ref{fig-wwcar-oc} indicates long-term 
changes in the pulsation period. The wave-like nature of these 
variations could be due to the light-time effect in a binary system. 
The available five radial velocity data covering only a week of 
observations \citep{Petal94} are, however, insufficient to study 
this possibility. The short-term scatter of the data points in 
Fig.~\ref{fig-wwcar-oc} reflects the observational error and 
uncertainties in the data analysis.\\

From a spectroscopic point of view, WW~Car has been a neglected
star. Even the F0 spectral type given by the SIMBAD database might be
incorrect for a genuine Cepheid variable. In addition to the five radial 
velocity measurements obtained in 1983 \citep{Petal94}, the atmospheric 
composition was studied by \citet{Letal11}, \citet{LL11}, and 
\citet{Uetal11}. 

\begin{table}
\caption{$O-C$ values of WW~Car (see the description in the text).}
\begin{tabular}{l@{\hskip2mm}r@{\hskip2mm}r@{\hskip2mm}c@{\hskip2mm}l}
\hline
JD$_{\odot}$ & $E\ $ & $O-C$ & $W$ & Data source\\
2\,400\,000 + &&&\\
\hline
23891.5175 & $-$6234 &  0.0284 & 1 & \citet{Sz26}\\
24285.4022 & $-$6150 &  0.0604 & 1 & \citet{Sz26}\\
34672.5746 & $-$3929 &  0.0200 & 1 & \citet{Wetal58}\\
35238.4720 & $-$3808 &  0.0224 & 2 & \citet{I61}\\
41047.0296 & $-$2566 & $-$0.0279 & 3 & \citet{P76}\\
44372.2203 & $-$1855 & $-$0.0548 & 3 & \citet{B08}\\
48249.3296 & $-$1026 & $-$0.0276 & 3 & {\it Hipparcos} \citep{ESA97}\\
48717.0190 &  $-$926 & $-$0.0200 & 3 & {\it Hipparcos} \citep{ESA97}\\
50578.3901 &  $-$528 & $-$0.0225 & 3 & \citet{B08}\\
51948.7159 &  $-$235 & $-$0.0044 & 3 & ASAS \citep{P02}\\
52369.6325 &  $-$145 & $-$0.0014 & 3 & ASAS \citep{P02}\\
52729.7555 &   $-$68 &  0.0066 & 3 & ASAS \citep{P02}\\
53047.7826 &     0 &  0.0101 & 3 & ASAS \citep{P02}\\
53150.6862 &    22 &  0.0237 & 3 & {\it INTEGRAL} OMC\\
53454.6642 &    87 &  0.0085 & 3 & ASAS \citep{P02}\\
53796.0799 &   160 &  0.0165 & 3 & ASAS \citep{P02}\\
54202.9568 &   247 &  0.0103 & 3 & ASAS \citep{P02}\\
54212.3011 &   249 &  0.0009 & 1 & {\it INTEGRAL} OMC\\
54535.0003 &   318 & $-$0.0003 & 3 & ASAS \citep{P02}\\
54876.4254 &   391 &  0.0171 & 3 & ASAS \citep{P02}\\
54918.5119 &   400 &  0.0122 & 3 & {\it INTEGRAL} OMC\\
55358.1264 &   494 &  0.0058 & 3 & AAVSO\\
\hline
\end{tabular}
\label{tab-wwcar-oc}
\end{table}

\subsection{FN~Velorum}
\label{sub-fnvel}

\begin{figure}
{\includegraphics[height=40mm, angle=0]{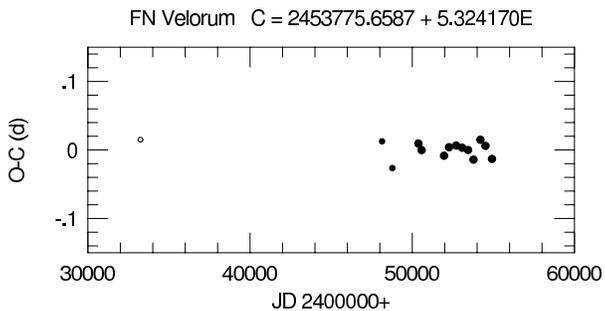}}
\caption[]{
$O-C$ diagram for FN~Vel.}
\label{fig-fnvel-oc}
\end{figure}

The brightness variability of FN~Velorum was revealed by 
\citet{OC51}. He already classified this variable as a Cepheid 
and published a correct period. Then this Cepheid had been 
neglected for decades. More recent photometric data are available 
from the Hipparcos mission \citep{ESA97}, the ASAS sky survey 
\citep{P02}, and the database containing Berdnikov's photometric 
observations of Cepheids \citep{B08}. The seasonal normal maxima 
listed in Table~\ref{tab-fnvel-oc} have been determined from these
data. The $O-C$ diagram is plotted in Fig.~\ref{fig-fnvel-oc}.
The plot can be approximated by a constant period corresponding
to the ephemeris \ref{fnvel-ephemeris} for the moments of the
maximum brightness. The scatter of the points in 
Fig.~\ref{fig-fnvel-oc} reflects the observational error and 
uncertainties in the analysis of the data.\\

\begin{table}
\caption{$O-C$ values of FN~Vel (see the description in the text).}
\begin{tabular}{l@{\hskip2mm}r@{\hskip2mm}r@{\hskip2mm}c@{\hskip2mm}l}
\hline
JD$_{\odot}$ & $E\ $ & $O-C$ & $W$ & Data source\\
2\,400\,000 + &&&\\
\hline
33240.35   &$-$3857 & 0.015 & 1 & \citet{OC51}\\
48137.3753 &$-$1059 & 0.0126 & 2 & {\it Hipparcos} \citep{ESA97}\\
48770.9125 &$-$940 & $-$0.0264 & 2 & {\it Hipparcos} \citep{ESA97}\\
50378.8476 &$-$638 & 0.0094 & 3 & \citet{B08}\\
50575.8323 &$-$601 &$-$0.0002 & 3 & \citet{B08}\\
51960.1083 &$-$341 &$-$0.0084 & 3 & ASAS \citep{P02}\\
52268.9226 &$-$283 & 0.0040 & 3 & ASAS \citep{P02}\\
52705.5070 &$-$201 & 0.0065 & 3 & ASAS \citep{P02}\\
53078.1957 &$-$131 & 0.0033 & 3 & ASAS \citep{P02}\\
53440.2359 & $-$63 &$-$0.0001 & 3 & ASAS \citep{P02}\\
53775.6446 &     0 &$-$0.0141 & 3 & ASAS \citep{P02}\\
54201.6072 &    80 & 0.0149 & 3 & ASAS \citep{P02}\\
54515.7243 &   139 & 0.0060 & 3 & ASAS \citep{P02}\\
54915.0180 &   214 &$-$0.0131 & 3 & ASAS \citep{P02}\\
\hline
\end{tabular}
\label{tab-fnvel-oc}
\end{table}

Spectroscopic observations of FN~Vel started only recently.
\citet{Letal11} studied the atmospheric chemical composition and
determined [Fe/H]\,=\,+0.06. Moreover, \citet{A13} revealed
that FN~Vel belongs to a spectroscopic binary system, and even
the spectroscopic orbit could be successfully determined from
his own extensive radial velocity measurement series. The
orbital period of the binary system is 471.654 days. This
spectroscopic binarity is a strong evidence of the reliability
of the method for binarity detection applied in the present paper.

\section{Conclusion}
\label{conclusion}

We used the so-called ``calcium-line method'' to investigate 
the presence of hot blue companions to 103 southern Cepheids.
In this method, the strong Balmer line, H$\epsilon$, of the 
companion superimposes on the Ca\,II\,H line of the Cepheid, 
resulting in the strengthening of the Ca\,II\,H line with respect 
to the Ca\,II\,K line in the compound spectrum of the binary system. 
(The Ca\,II\,H \& K lines have practically equal depths in single 
Cepheids).

The method allowed us to recover eight Cepheids with known blue 
companions in our sample and led to the discovery of hot companions 
for two more Cepheids, WW~Car and FN~Vel. In the case of FN~Vel,
this is an independent confirmation of binarity published by
\citet{A13} in his PhD Thesis. WW~Car has also been suspected in 
having a blue companion from photometry \citep{MF80}.

As Cepheids are used as standard candles in determining the 
cosmic distance scale, it is important to disentangle 
the luminosity of the Cepheids from that of their companion 
when calibrating the $P$-$L$ relationship. Therefore
the binarity of Cepheids should be studied on a star-by-star basis.

\section*{Acknowledgements}

FCh acknowledges SCOPES grant No. IZ73Z0-152485 for financial support.
LSz was supported by the ESTEC Contract no.\,4000106398/12/NL/KML.
The {\it INTEGRAL\/} photometric data, pre-processed by ISDC, 
have been retrieved from the OMC Archive at CAB (INTA-CSIC).
The observer of the AAVSO photometric data was Neil Butterworth.
Authors thank the anonymous refree for her/his valuable comments 
that significantly improved the paper.

\label{lastpage}

\bsp

\end{document}